\documentclass{article}

\usepackage[final,nonatbib]{neurips_2022_ml4ps}

\usepackage{float}
\usepackage{chngpage}
\usepackage{xcolor}

\PassOptionsToPackage{numeric,compress}{natbib}
\usepackage[utf8]{inputenc} % allow utf-8 input
\usepackage[T1]{fontenc}    % use 8-bit T1 fonts
\usepackage{hyperref}       % hyperlinks
\hypersetup{
    colorlinks=true,
    linkcolor=black,
    filecolor=magenta,      
    urlcolor=blue,
    citecolor=blue,
}

\usepackage{url}            % simple URL typesetting
\usepackage{booktabs}       % professional-quality tables
\usepackage{amsfonts}       % blackboard math symbols
\usepackage{nicefrac}       % compact symbols for 1/2, etc.
\usepackage{microtype}      % microtypography
\usepackage{lipsum}
\usepackage{graphicx}
\usepackage{amsmath}
\usepackage{multirow}
\usepackage{xspace}
\usepackage{cite}
\usepackage{listings}

\usepackage{color}
\definecolor{deepblue}{rgb}{0,0,0.5}
\definecolor{deepred}{rgb}{0.6,0,0}
\definecolor{deepgreen}{rgb}{0,0.5,0}

\usepackage{subcaption}
\usepackage{array}
\newcolumntype{-}{>{\global\let\currentrowstyle\relax}}
\newcolumntype{^}{>{\currentrowstyle}}

\DeclareFixedFont{\ttb}{T1}{txtt}{bx}{n}{10} % for bold
\DeclareFixedFont{\ttm}{T1}{txtt}{m}{n}{10}  % for normal

\newcommand\pythonstyle{\lstset{
language=Python,
basicstyle=\ttm,
otherkeywords={def},             % Add keywords here
keywordstyle=\ttb\color{deepblue},
emph={MyClass,__init__},          % Custom highlighting
emphstyle=\ttb\color{deepred},    % Custom highlighting style
stringstyle=\color{deepgreen},
frame=tb,                         % Any extra options here
showstringspaces=false            % 
}}

\lstnewenvironment{python}[1][]
{
\pythonstyle
\lstset{#1}
}
{}

\newcommand\pythoninline[1]{{\pythonstyle\lstinline!#1!}}

\title{Do graph neural networks learn traditional jet substructure?}

\newcommand{\kt}{\ensuremath{k_{\mathrm{T}}}\xspace}
\newcommand{\pt}{\ensuremath{p_{\mathrm{T}}}\xspace}
\newcommand{\GeV}{\ensuremath{\,\text{Ge\hspace{-.08em}V}}\xspace}
\newcommand{\TeV}{\ensuremath{\,\text{Te\hspace{-.08em}V}}\xspace}

\newcommand{\PYTHIA} {{\textsc{pythia}}\xspace}

\newcommand{\DELPHES} {{\textsc{delphes}}\xspace}
\newcommand{\FastJet} {{\textsc{fastjet}}\xspace}

\begin{document}

\author{
    Farouk Mokhtar, Raghav Kansal, Javier Duarte\\
    University of California San Diego \\
    La Jolla, CA 92093, USA\\
    \texttt{\{fmokhtar,rkansal,jduarte\}@ucsd.edu}
}

\begin{flushright}
FERMILAB-CONF-22-889-CMS-PPD
\end{flushright}
\maketitle

\begin{abstract}
    At the CERN LHC, the task of jet tagging, whose goal is to infer the origin of a jet given a set of final-state particles, is dominated by machine learning methods.
    Graph neural networks have been used to address this task by treating jets as point clouds with underlying, learnable, edge connections between the particles inside.
    We explore the decision-making process for one such state-of-the-art network, ParticleNet, by looking for relevant edge connections identified using the layerwise-relevance propagation technique.
    As the model is trained, we observe changes in the distribution of relevant edges connecting different intermediate clusters of particles, known as subjets.
    The resulting distribution of subjet connections is different for signal jets originating from top quarks, whose subjets typically correspond to its three decay products, and background jets originating from lighter quarks and gluons.
    This behavior indicates that the model is using traditional jet substructure observables, such as the number of \emph{prongs}---energetic particle clusters---within a jet, when identifying jets.
\end{abstract}

\paragraph{Introduction}

The CERN LHC is an enormous data generation machine: producing about one petabyte of sensor-level collision data per second~\cite{Butler:2013kka}.
This collision data passes a series of online (and offline) filtering through complex trigger systems optimized for data reduction to achieve reasonable storage capacity for further analyses.
Machine learning (ML) algorithms are implemented throughout the entire LHC workflow, from the early steps of data reduction and compression at the level-1 trigger for online filtering~\cite{Duarte:2018ite,CERN-LHCC-2020-004,Deiana:2021niw} to full-scale physics analyses~\cite{Guest:2018yhq}. 
In particular, graph neural networks (GNNs) are found to be performant for a wide variety of tasks at the LHC~\cite{Shlomi:2020gdn,Duarte:2020ngm} since collider data can be optimally described, in many cases (for instance when it comes to jets), as a point cloud. 
Jets are showers of particles initiated by quarks and gluons, and jet substructure~\cite{Larkoski:2017jix,Kogler:2018hem} studies the radiation pattern of those jets. 
A particular observable of interest when identifying jets is the number of \emph{prongs}---energetic clusters of particles---within a jet. 
For example, it is common for a hadronically decaying top quark jet to exhibit a three-prong substructure.

Examples of jet tasks that can be addressed with GNNs include jet simulation~\cite{kansal2021particle}, jet clustering~\cite{Qasim:2019otl,Kieseler:2020wcq,Guo:2020vvt}, jet mass regression~\cite{CMS-DP-2021-017}, and jet classification (or jet tagging)~\cite{Moreno:2019bmu,Moreno:2019neq,Qu:2019gqs,Mikuni:2020wpr}, which is the subject of this paper.
ParticleNet~\cite{Qu:2019gqs} is a state-of-the-art GNN developed to address the problem of jet tagging by treating jets as point clouds with underlying, learnable, edge connections between the particles.

ML algorithms are often referred to as black boxes because their behavior is not easily interpretable. 
A collection of techniques, called explainable artificial intelligence (XAI)~\cite{XAI}, attempt to explain the decision-making of ML models. 
One such technique is layerwise relevance propagation (LRP)~\cite{LRP,Montavon2019}, which relies on computing and assigning relevancy ($R$) scores to neurons of an ML model to measure their influence on the prediction.
Explaining ParticleNet with LRP is interesting as we can assign $R$ scores to the learned edges connecting the particle nodes in the point cloud. 
These \emph{edge $R$ scores} highlight particular connections between particles that were found to be the most influential on the final output, as the model learns to discriminate jets.
This helps us understand the extent to which the model is utilizing the physics we know, for example the number of prongs, when identifying jets.

\paragraph{Related Work}
\label{sec:related}

ML models implemented at the LHC tend to lack physics-informed inductive bias that are built into alternative algorithms, which makes interpretability studies on such models especially important~\cite{Agarwal:2020fpt,Lai:2020byl,Neubauer:2022zpn}.
Ref.~\cite{Mokhtar:2021bkf} attempts to explain the machine-learned particle-flow~\cite{Pata:2021oez,Pata:2022wam} algorithm, which is a GNN developed for the task of particle-flow reconstruction in CMS. 
We build upon this work to extend the LRP technique for ParticleNet, which makes use of specific graph convolutions.

At its core, the LRP technique relies on simple operations that systematically redistribute the output prediction of an ML model over neurons in each layer. 
The result of the redistribution is an $R$ score per neuron that reflects its importance to the overall prediction.
GNNs naturally impose a challenge to the application of LRP due to their inherent complexity. 
The attempts to adapt LRP for GNNs include the GNN-LRP ~\cite{GNNLRP} and GLRP methods~\cite{GLRP}. 
Since each GNN model realizes different, and possibly unique, operations, it is hard to find an implementation that fits all.

\paragraph{LRP applied to ParticleNet}
\label{sec:Modifications}

We train a ParticleNet model on the benchmark top quark jet tagging dataset~\cite{Kasieczka:2019dbj} found on the Zenodo platform ~\cite{kasieczka_gregor_2019_2603256} under the CC BY 4.0 license. 
Our code is also publicly available under the MIT license~\cite{xai4hep}.
We choose top quark jets as our signal in particular because such jets are known to have distinct features.
For example, we investigate whether the model makes use of the three-pronged substructure of top quark jets to distinguish them from lighter quark or gluons jets (collectively referred to as quantum chromodynamic (QCD) jets), which are typically single-pronged~\cite{Thaler:2010tr,Larkoski:2017jix,Kogler:2018hem}. 
% More information on the dataset can be found in \autoref{app:performance}.

The dataset is composed of jets from 14\TeV collisions, corresponding either to hadronically decaying top quarks for signal or QCD dijets for background.
The dataset is generated using \PYTHIA8~\cite{Sjostrand:2014zea} and simulated using the \DELPHES~\cite{deFavereau:2013fsa} without multiple parton interactions and without pileup.
Particle-flow entries are clustered using the anti-\kt algorithm~\cite{Cacciari:2008gp} implemented in the \FastJet package~\cite{Cacciari:2011ma} with a radius parameter of $R=0.8$.
Jets with transverse momentum (\pt) between 550 and 650\GeV are selected.
All top quark jets are matched to a parton-level top quark within a distance of $\Delta R=0.8$, and to all top quark decay partons within $\Delta R =0.8$.
Jets are required to have $|\eta| < 2$.
The 200 \pt-leading jet constituent four-momenta are accessible, sorted by decreasing \pt, and are fed as input to our ParticleNet model. 
We build the model in PyTorch Geometric (PyG)~\cite{PyTorchGeometric} which has the advantage of handling variable sized graphs. 
This means that our input jets can have an arbitrary number of jet constituents.
For example, a jet with 60-constituents will correspond to an input with dimensions $[60, 4]$, containing the four-momenta $(p_x, p_y, p_z, E)$ of each constituent.

The ParticleNet model primarily makes use of the EdgeConv block consisting of dynamic graph convolutions proposed in Ref.~\cite{DGCNN} to learn the graph structure.
Following the same choice proposed in the ParticleNet paper, we choose to have three EdgeConv blocks in our model.
The EdgeConv block builds a graph using the $k$-nearest neighbor (kNN) technique.
The model then invokes a fully-connected deep neural network to learn edge weights to be assigned to each edge connection.
This is followed by a message-passing step where node features are updated using a pooling operation on the learned edge weights, followed by a pooling operation on all nodes, and lastly, a final fully-connected neural network layer.
The model outputs a single number, followed by a sigmoid activation function.

\begin{figure}[htpb]
    \centering
    \includegraphics[width=0.6\textwidth]{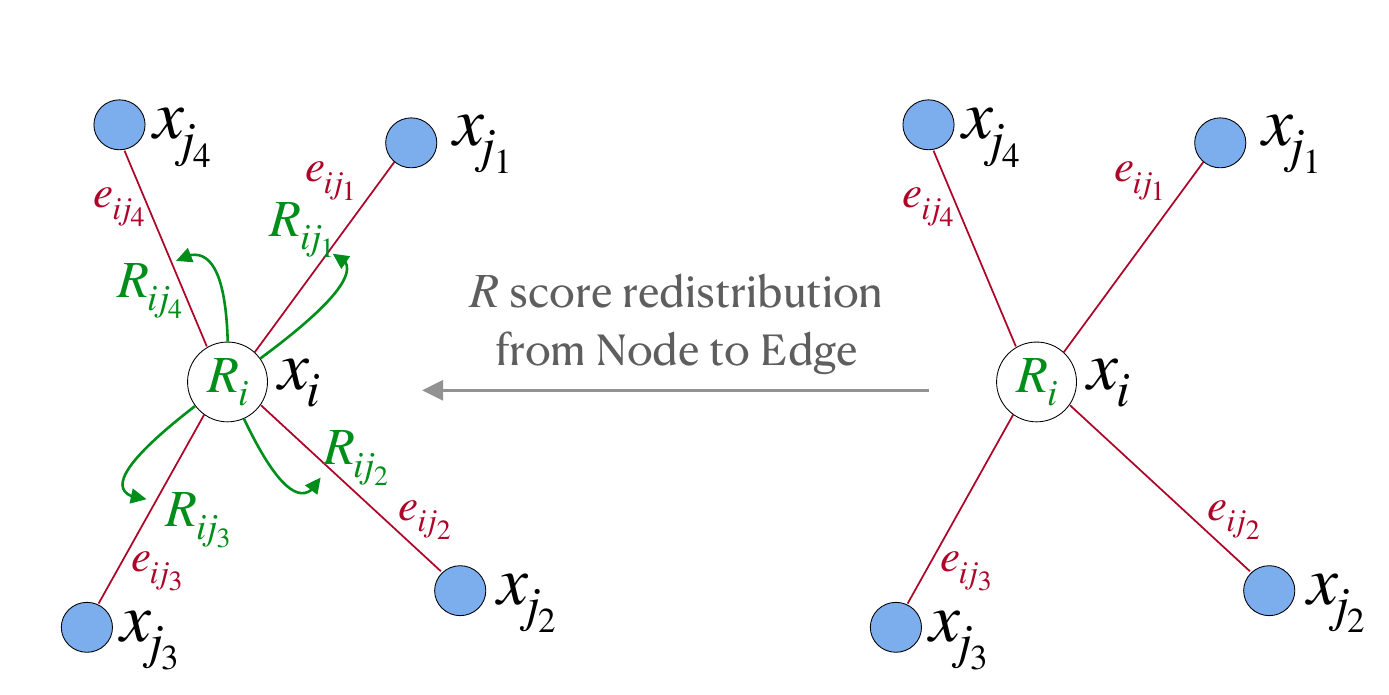}
    \caption{
        Redistribution of the $R$ scores of node $x_i$ over the edges connecting the node with its ${K}$ neighbors labeled by $x_{j_k}$ to obtain the edge $R$ scores $R_{ij_k}$.
    }
    \label{fig:rscores_redistribution}
\end{figure}

From the LRP standpoint, the operations involved in the EdgeConv block can all be cast as matrix operations that imitate a basic multilayer perceptron operation.
This makes it straightforward to use the standard LRP rules after some minimal manipulations.
In \autoref{fig:rscores_redistribution} we illustrate how the redistribution of $R$ scores occurs from a single node to the surrounding edges to obtain the edge $R$ scores.
The edge $R$ score $R_{ij_k}$ for a given edge $e_{ij_k}$ connecting node $x_i$ with its $k$th neighboring node $x_{j_k}$ is computed using the formula:
\begin{equation}\label{eq:LRP}
    R_{ij_k} = \frac{e_{ij_k}}{\sum_{k'}^{K}e_{ij_k'}}R_i
\end{equation}
where ${K}$ denotes the number of neighbors of node $x_i$.

\paragraph{Results}
\label{sec:Results}

We train a ParticleNet model for further LRP tests (see \autoref{app:performance} for details on the training parameters). 
We run the LRP evaluation on 10,000 jet samples that were reserved for testing during the ParticleNet model training.
The evaluation takes 2.5 hours on an Apple M1 Pro chip.
In addition, we also run the LRP code on 10 randomly initialized untrained models for baseline comparison with the trained model.
For each jet, we propagate the $R$ score of the output node backwards, and are able to obtain an intermediate relevancy-graph for each graph constructed in an EdgeConv block. 
We refer to these as ``edge $R$ graphs''.

\paragraph[Edge R graphs]{Edge $R$ graphs}
The edge $R$ graph represents the jet constituents as nodes in $(\eta, \phi)$ space and interconnections as edges, whose intensities correspond to the respective edge $R$ score.
Each node's intensity corresponds to the relative \pt of the corresponding particle.
In principle, we can propagate $R$ scores backwards until we reach the input layer, but we find this unnecessary because we focus on the GNN's learned graph connectivity.
Edge $R$ graphs can be constructed for each EdgeConv block.
Note that every block constructs a new graph using the kNN technique, so we expect the adjacency matrix to change with each block.

% For a sample of correctly classified top quark jets, we observe that with every EdgeConv block the model is attempting to connect nodes that are further apart. \autoref{fig:results} shows Edge-Rgraphs of the three EdgeConv blocks for a correctly classified top quark jet. 
% Furthermore, we often observe these triangles of three high relevant edges connecting, what seems to us, different subclusters (or subjets). 

\begin{figure}[htpb]
    \centering
    \includegraphics[width=1\textwidth]{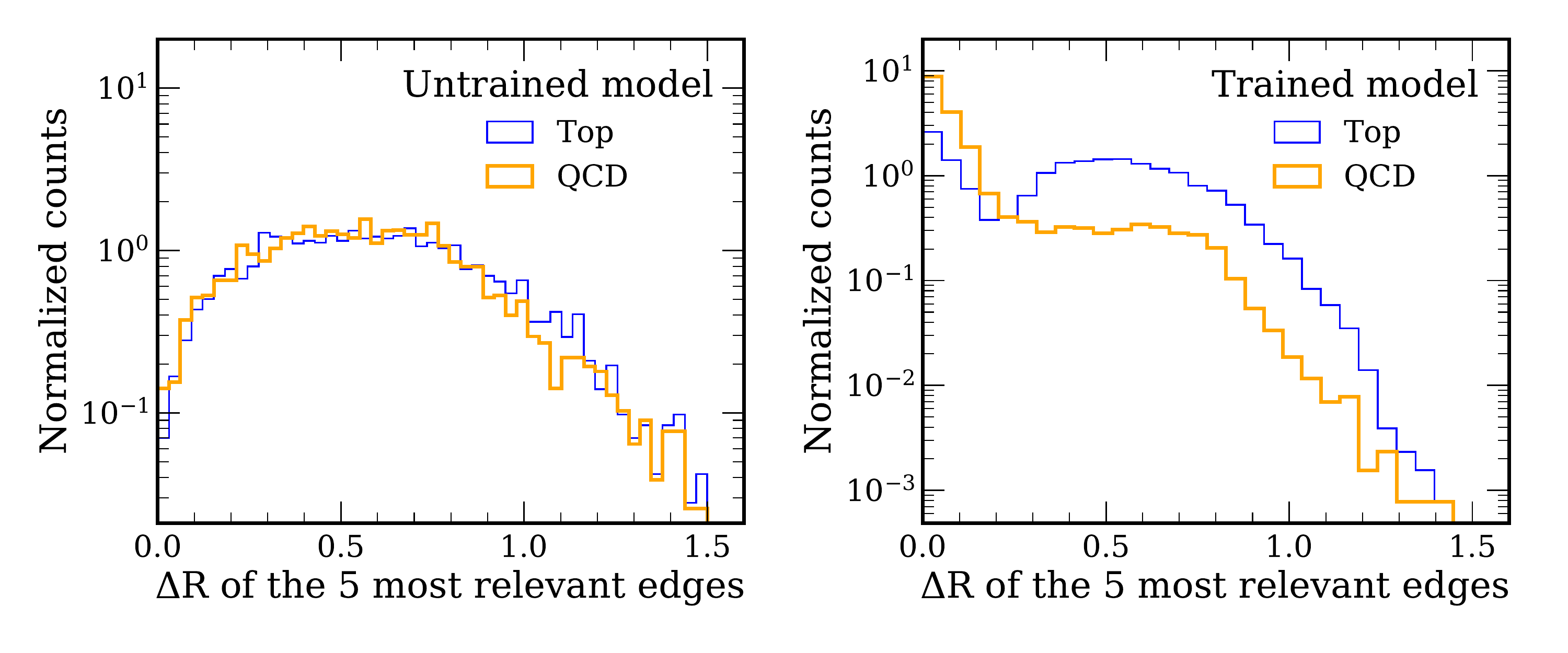}
    \caption{
        We present the distribution of the $\Delta R$ of the five most relevant edges for top quark jets (blue) versus QCD jets (orange) for an untrained ParticleNet model (left) and the learned distribution by a trained ParticleNet model (right).
        The result for the untrained model is an average over 10 randomly initialized models.}
    \label{fig:deltaR}
\end{figure}

Looking at sample edge $R$ graphs, we observe that by the last EdgeConv block, the model is learning to connect nodes farther apart in $(\eta, \phi)$ space.
This happens more often for top quark than for QCD jets, and such connections have high edge relevancy scores. We present sample edge $R$ graphs in \autoref{app:sample_graphs} for both top quark and QCD jets.
We aggregate over many jets by exploring how the distribution of the $\Delta R$ of the five most relevant edges in the last EdgeConv block changes as the model trains.
From \autoref{fig:deltaR}, we see that the model learns to emphasize higher $\Delta R$ edges for top quark jets compared to QCD jets. 
The results shown for the randomly initialized untrained models act as a baseline to compare with the trained model~\cite{sanity_saliency}.
If the LRP method depends on the learned parameters of the model, we expect its output to differ substantially between the two cases, indicating what the model has learned.
Moreover, since the top quark and QCD distributions for an untrained model are nearly identical, there is nothing intrinsic in the dataset that biases the model toward connecting differently-separated edges between the two classes.

This raises the question of whether the model is learning jet substructure, and whether it is using observables such as the number of prongs within a jet, when identifying top quark jets.
We attempt to answer this question by looking at the fraction of relevant edges that connect particles from different intermediate clusters, or \emph{subjets}.
We use the Cambridge--Aachen (CA)  algorithm~\cite{Dokshitzer:1997in,Wobisch:1998wt} to decluster the jets into exactly three subjets.
We choose the CA algorithm because the metric it uses to cluster particles relies purely on their spatial positions and not their momenta.
It has also been used in previous studies of top quark jets~\cite{CMS-PAS-JME-09-001}.
We use the \FastJet Python package to first recluster each jet inclusively using the CA algorithm with a radius parameter of $R=0.8$.
Then, we trace backward through the history of the sequential combinations until we find exactly three subjets, i.e. the final subjets to be combined by the CA algorithm.

\begin{figure}[htpb]
    % \centering
    \includegraphics[width=1\textwidth]{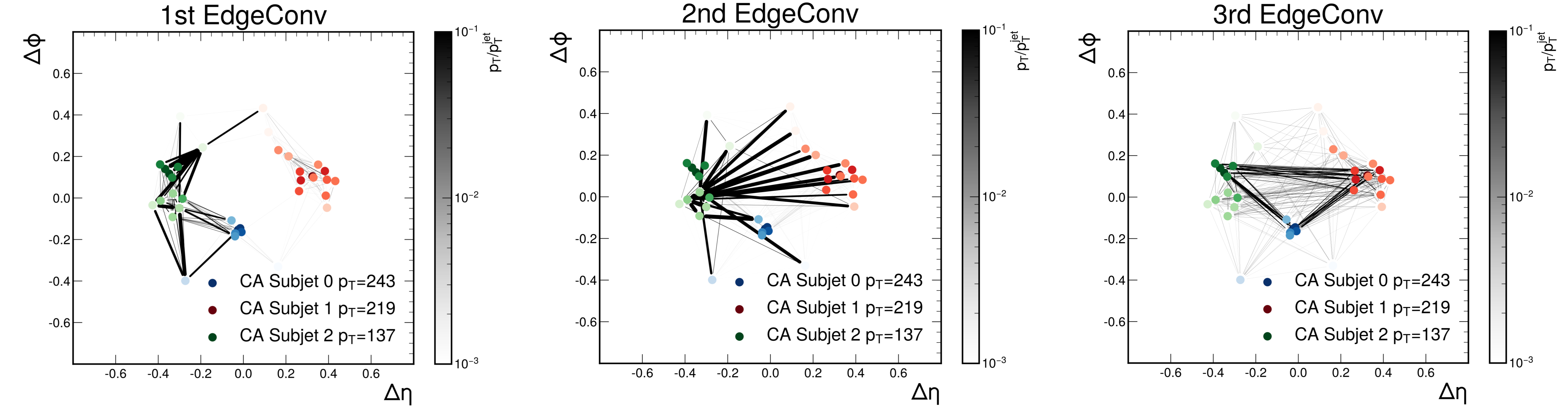}    
    \caption{
        The three edge $R$ graphs for a true top quark jet corresponding to the three graphs learned with each EdgeConv block. 
        The jet constituents are represented as nodes in $(\eta, \phi)$ space with interconnections as edges, whose intensities correspond to the connection's edge $R$ score.
        Each node's intensity corresponds to the relative \pt of the corresponding particle.
        Constituents belonging to the three different CA subjets are shown in blue, red, and green in descending \pt order.
        We observe that by the last EdgeConv block the model learns to rely more on edge connections between the different subjets.
    }
    \label{fig:wow}
\end{figure}

In \autoref{fig:wow}, we observe that with each learned graph at each EdgeConv block, the model learns to rely more on edges connecting different subjets when discriminating top quark from QCD.
For the first EdgeConv block in \autoref{fig:wow} (left), only the nearest neighbors in $(\eta, \phi)$ space are connected, while for the third EdgeConv block in \autoref{fig:wow} (right) more long-distance connections between the different CA subjets are present.

\begin{figure}[htpb]
    \centering
    \includegraphics[width=1\textwidth]{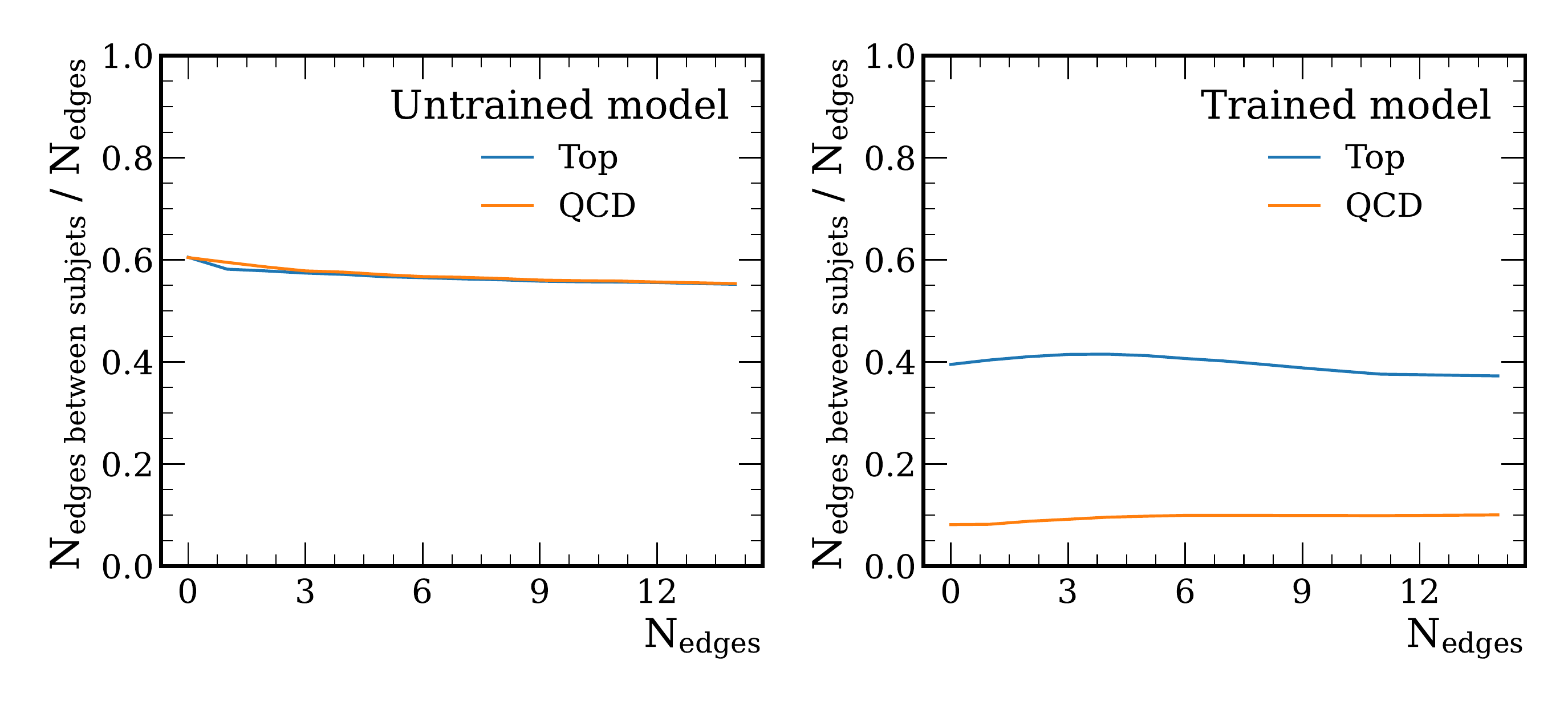}
    \caption{
        The fraction of edges connecting different subjets $N_{\text{edges between subjets}}$ among the $N$ most relevant edges $N_{\text{edges}}$, for different values of $N$. 
        Results are shown averaged over 10 randomly initialized untrained ParticleNet models (left) and for the trained ParticleNet model (right).
    }
    \label{fig:scaling_up}
\end{figure}

To aggregate over many jets and extract a statistically meaningful result, we explore the fraction of edges connecting different subjets $N_{\text{edges between subjets}}$ among the $N$ most relevant edges $N_{\text{edges}}$. 
We scan different values of $N$ and we test this for 10 untrained ParticleNet models, and a trained ParticleNet model.
The results are shown in \autoref{fig:scaling_up}.

From \autoref{fig:scaling_up}, we can see that an untrained model treats both top quark and QCD jets the same, however, as it trains, it learns to connect and rely more on edges connecting different subjets when it is identifying top quark jets compared to QCD.
This indicates that the model is learning the difference in the respective substructure, and is learning to use observables such as the number of prongs, when identifying jets.

\vspace{-1ex}

\paragraph{Summary}
\label{sec:summary}
We attempt to explain the decision-making process of the ParticleNet model for top quark jet tagging, and find that the model learns different graph connectivity for discriminating top quark jets from QCD background.
We observe that, on average, the most relevant edges learned for top quark jets have larger $\Delta R$ separation than for QCD jets. 
This motivates us to further explore the distribution of relevant edges connecting different subjets. 
We find that as the model trains, it learns to rely more on edges connecting different subjets when classifying top quark jets compared to QCD jets, indicating that the model is learning the difference in the respective substructure, and is learning to use observables such as the number of prongs, when identifying jets.
This simple result motivates us to further explore how the relevant edges correspond to known jet substructure variables, like $N$-subjettiness~\cite{Thaler:2010tr}, energy correlation functions~\cite{Larkoski:2013eya,Moult:2016cvt}, and energy flow polynomials~\cite{Komiske:2017aww}.

\vspace{-1ex}

\paragraph{Limitations}
One potential limitation is to what degree the results shown in this work depend on the chosen subjet finding algorithm.
Another matter of concern is the robustness of the model, since LRP (and most XAI techniques) rely on sensitive changes to the input and their effect on the output. 
Works such as Ref.~\cite{interpretability_and_robustness} have previously shown that it is mathematically impossible to have both attribution-based explanations (that provide recourse) and robustness at the same time. 
Nevertheless, it is still valuable to explore the interpretability of widely used models.

% Since the LRP technique highlights the most relevant features of the input, and in case of this paper, the relevant edges/connections, this can potentially give us insights on intelligently reducing the input size of our data, or the number of nearest neighbors chosen without suffering significant performance penalty. This could be helpful, in terms of speed up and lower storage since the planned LHC upgrade would increase the size of our data by an expected factor of 10.

\paragraph{Broader Impact}
\label{sec:impacts}
Attempts to explain and interpret ML algorithms are useful in general to increase confidence in the model prediction as well as to gain insights on making models more performant and robust. 
Interpretability studies are especially important for ML models implemented at the LHC because such models tend to lack physics-informed inductive bias that are built into alternative algorithms. 
This effort explores the application of an XAI method on a state-of-the-art GNN model developed for jet tagging at the LHC, and attempts to show that the model is indeed learning useful physics that is already being exploited in physics-designed alternatives.
This opens the door for potentially extracting information and relationships exploited by the ML model that may not be fully utilized by such alternative algorithms.

\begin{ack}
    F.~M. is supported by an Hal{\i}c{\i}o\u{g}lu Data Science Institute (HDSI) fellowship. 
    % and an Institute for Research and Innovation in Software for High Energy Physics (IRIS-HEP) fellowship through the U.S. National Science Foundation (NSF) under Cooperative Agreement OAC-1836650.
    F.~M., R.~K., and J.~D. are supported by the US Department of Energy (DOE), Office of Science, Office of High Energy Physics Early Career Research program under Award No. DE-SC0021187, the DOE, Office of Advanced Scientific Computing Research under Award No. DE-SC0021396 (FAIR4HEP), and the NSF HDR Institute for Accelerating AI Algorithms for Data Driven Discovery (A3D3) under Cooperative Agreement OAC-2117997.
    R.~K. is also supported by the LHC Physics Center at Fermi National Accelerator Laboratory, managed and operated by Fermi Research Alliance, LLC under Contract No. DE-AC02-07CH11359 with the DOE.
    This work was performed using the Pacific Research Platform Nautilus HyperCluster supported by NSF awards CNS-1730158, ACI-1540112, ACI-1541349, OAC-1826967, the University of California Office of the President, and the University of California San Diego's California Institute for Telecommunications and Information Technology/Qualcomm Institute.
    Thanks to CENIC for the 100\,Gpbs networks.
    Funding for cloud credits was supported by NSF Award \#1904444 Internet2 Exploring Clouds to Accelerate Science (E-CAS).
\end{ack}

\bibliographystyle{cms_unsrt}
\bibliography{bibliography}

\providecommand{\href}[2]{#2}\begingroup\raggedright\begin{thebibliography}{10}%
\makeatletter
\providecommand{\hrefCMSnoop }[0]{\@secondoftwo}%
\makeatother
\providecommand{\doi}{\texttt{doi:}\begingroup \urlstyle{tt}\Url}

\bibitem{Butler:2013kka}
\hrefCMSnoop {}{M.~Butler, R.~Mount, and M.~Hildreth, ``{Working Group Report:
  Storage and Data Management}'',} in \textit{ {Community Summer Study 2013}:
  {Snowmass on the Mississippi}}.
\newblock 2013.
\newblock \href{http://www.arXiv.org/abs/1311.4580}{\texttt{arXiv:1311.4580}}.

\bibitem{Duarte:2018ite}
\hrefCMSnoop {}{J.~Duarte {et~al.}, ``{Fast inference of deep neural networks
  in FPGAs for particle physics}'',} \textit{ J. Instrum.} \textbf{ 13} (2018)
  P07027,
  \href{http://dx.doi.org/10.1088/1748-0221/13/07/P07027}{\doi{10.1088/1748-0221/13/07/P07027}},
  \href{http://www.arXiv.org/abs/1804.06913}{\texttt{arXiv:1804.06913}}.

\bibitem{CERN-LHCC-2020-004}
\href {https://cds.cern.ch/record/2714892}{{CMS} Collaboration, ``The {Phase-2}
  upgrade of the {CMS} {Level-1} trigger'',} CMS Technical Design Report
  CERN-LHCC-2020-004. CMS-TDR-021, 2020.

\bibitem{Deiana:2021niw}
\hrefCMSnoop {}{A.~M. Deiana {et~al.}, ``{Applications and Techniques for Fast
  Machine Learning in Science}'',} \textit{ Front. Big Data} \textbf{ 5} (2022)
  787421,
  \href{http://dx.doi.org/10.3389/fdata.2022.787421}{\doi{10.3389/fdata.2022.787421}},
  \href{http://www.arXiv.org/abs/2110.13041}{\texttt{arXiv:2110.13041}}.

\bibitem{Guest:2018yhq}
\hrefCMSnoop {}{D.~Guest, K.~Cranmer, and D.~Whiteson, ``{Deep Learning and its
  Application to LHC Physics}'',} \textit{ Ann. Rev. Nucl. Part. Sci.} \textbf{
  68} (2018) 161--181,
  \href{http://dx.doi.org/10.1146/annurev-nucl-101917-021019}{\doi{10.1146/annurev-nucl-101917-021019}},
  \href{http://www.arXiv.org/abs/1806.11484}{\texttt{arXiv:1806.11484}}.

\bibitem{Shlomi:2020gdn}
\hrefCMSnoop {}{J.~Shlomi, P.~Battaglia, and J.-R. Vlimant, ``{Graph Neural
  Networks in Particle Physics}'',} \textit{ Mach. Learn.: Sci. Technol.}
  \textbf{ 2} (2021) 021001,
  \href{http://dx.doi.org/10.1088/2632-2153/abbf9a}{\doi{10.1088/2632-2153/abbf9a}},
  \href{http://www.arXiv.org/abs/2007.13681}{\texttt{arXiv:2007.13681}}.

\bibitem{Duarte:2020ngm}
\hrefCMSnoop {}{J.~Duarte and J.-R. Vlimant, ``Graph neural networks for
  particle tracking and reconstruction'',} in \textit{ Artificial Intelligence
  for Particle Physics}.
\newblock World Scientific Publishing, 2020.
\newblock
  \href{http://www.arXiv.org/abs/2012.01249}{\texttt{arXiv:2012.01249}}.
\newblock Submitted to \emph{Int. J. Mod. Phys. A}.
  \href{http://dx.doi.org/10.1142/12200}{\doi{10.1142/12200}}.

\bibitem{Larkoski:2017jix}
\hrefCMSnoop {}{A.~J. Larkoski, I.~Moult, and B.~Nachman, ``{Jet Substructure
  at the Large Hadron Collider: A Review of Recent Advances in Theory and
  Machine Learning}'',} \textit{ Phys. Rept.} \textbf{ 841} (2020) 1,
  \href{http://dx.doi.org/10.1016/j.physrep.2019.11.001}{\doi{10.1016/j.physrep.2019.11.001}},
  \href{http://www.arXiv.org/abs/1709.04464}{\texttt{arXiv:1709.04464}}.

\bibitem{Kogler:2018hem}
\hrefCMSnoop {}{R.~Kogler {et~al.}, ``{Jet Substructure at the Large Hadron
  Collider: Experimental Review}'',} \textit{ Rev. Mod. Phys.} \textbf{ 91}
  (2019), no.~4, 045003,
  \href{http://dx.doi.org/10.1103/RevModPhys.91.045003}{\doi{10.1103/RevModPhys.91.045003}},
  \href{http://www.arXiv.org/abs/1803.06991}{\texttt{arXiv:1803.06991}}.

\bibitem{kansal2021particle}
R.~Kansal\href
  {https://papers.nips.cc/paper/2021/hash/c8512d142a2d849725f31a9a7a361ab9-Abstract.html}{
  {et~al.}, ``Particle cloud generation with message passing generative
  adversarial networks'',} in \textit{ {Advances in Neural Information
  Processing Systems}}, volume~34.
\newblock Curran Associates, Inc., 2021.
\newblock
  \href{http://www.arXiv.org/abs/2106.11535}{\texttt{arXiv:2106.11535}}.

\bibitem{Qasim:2019otl}
\hrefCMSnoop {}{S.~R. Qasim, J.~Kieseler, Y.~Iiyama, and M.~Pierini, ``Learning
  representations of irregular particle-detector geometry with
  distance-weighted graph networks'',} \textit{ Eur. Phys. J. C} \textbf{ 79}
  (2019) 608,
  \href{http://dx.doi.org/10.1140/epjc/s10052-019-7113-9}{\doi{10.1140/epjc/s10052-019-7113-9}},
  \href{http://www.arXiv.org/abs/1902.07987}{\texttt{arXiv:1902.07987}}.

\bibitem{Kieseler:2020wcq}
\hrefCMSnoop {}{J.~Kieseler, ``{Object condensation: one-stage grid-free
  multi-object reconstruction in physics detectors, graph and image data}'',}
  \textit{ Eur. Phys. J. C} \textbf{ 80} (2020), no.~9, 886,
  \href{http://dx.doi.org/10.1140/epjc/s10052-020-08461-2}{\doi{10.1140/epjc/s10052-020-08461-2}},
  \href{http://www.arXiv.org/abs/2002.03605}{\texttt{arXiv:2002.03605}}.

\bibitem{Guo:2020vvt}
\hrefCMSnoop {}{J.~Guo, J.~Li, T.~Li, and R.~Zhang, ``{Boosted Higgs boson jet
  reconstruction via a graph neural network}'',} \textit{ Phys. Rev. D}
  \textbf{ 103} (2021), no.~11, 116025,
  \href{http://dx.doi.org/10.1103/PhysRevD.103.116025}{\doi{10.1103/PhysRevD.103.116025}},
  \href{http://www.arXiv.org/abs/2010.05464}{\texttt{arXiv:2010.05464}}.

\bibitem{CMS-DP-2021-017}
\href {https://cds.cern.ch/record/2777006}{{CMS} Collaboration, ``{Mass
  regression of highly-boosted jets using graph neural networks}'',} {CMS
  Detector Performance Note} CMS-DP-2021-017, 2021.

\bibitem{Moreno:2019bmu}
E.~A. Moreno\hrefCMSnoop {}{ {et~al.}, ``{JEDI-net: a jet identification
  algorithm based on interaction networks}'',} \textit{ Eur. Phys. J. C}
  \textbf{ 80} (2020) 58,
  \href{http://dx.doi.org/10.1140/epjc/s10052-020-7608-4}{\doi{10.1140/epjc/s10052-020-7608-4}},
  \href{http://www.arXiv.org/abs/1908.05318}{\texttt{arXiv:1908.05318}}.

\bibitem{Moreno:2019neq}
E.~A. Moreno\hrefCMSnoop {}{ {et~al.}, ``{Interaction networks for the
  identification of boosted $H \rightarrow b\overline{b}$ decays}'',} \textit{
  Phys. Rev. D} \textbf{ 102} (2020) 012010,
  \href{http://dx.doi.org/10.1103/PhysRevD.102.012010}{\doi{10.1103/PhysRevD.102.012010}},
  \href{http://www.arXiv.org/abs/1909.12285}{\texttt{arXiv:1909.12285}}.

\bibitem{Qu:2019gqs}
\hrefCMSnoop {}{H.~Qu and L.~Gouskos, ``{ParticleNet: Jet Tagging via Particle
  Clouds}'',} \textit{ Phys. Rev. D} \textbf{ 101} (2020) 056019,
  \href{http://dx.doi.org/10.1103/PhysRevD.101.056019}{\doi{10.1103/PhysRevD.101.056019}},
  \href{http://www.arXiv.org/abs/1902.08570}{\texttt{arXiv:1902.08570}}.

\bibitem{Mikuni:2020wpr}
\hrefCMSnoop {}{V.~Mikuni and F.~Canelli, ``{ABCNet: An attention-based method
  for particle tagging}'',} \textit{ Eur. Phys. J. Plus} \textbf{ 135} (2020),
  no.~6, 463,
  \href{http://dx.doi.org/10.1140/epjp/s13360-020-00497-3}{\doi{10.1140/epjp/s13360-020-00497-3}},
  \href{http://www.arXiv.org/abs/2001.05311}{\texttt{arXiv:2001.05311}}.

\bibitem{XAI}
W.~Samek\hrefCMSnoop {}{ {et~al.}, eds., ``Explainable {AI}: Interpreting,
  Explaining and Visualizing Deep Learning''}.
\newblock Springer International Publishing, Cham, Switzerland, (2019).
\newblock
  \href{http://dx.doi.org/10.1007/978-3-030-28954-6}{\doi{10.1007/978-3-030-28954-6}}.

\bibitem{LRP}
S.~Bach\hrefCMSnoop {}{ {et~al.}, ``On pixel-wise explanations for non-linear
  classifier decisions by layer-wise relevance propagation'',} \textit{ PLOS
  ONE} \textbf{ 10} (2015) e0130140,
  \href{http://dx.doi.org/10.1371/journal.pone.0130140}{\doi{10.1371/journal.pone.0130140}}.

\bibitem{Montavon2019}
G.~Montavon\hrefCMSnoop {}{ {et~al.}, ``Layer-wise relevance propagation: An
  overview'',} in Samek {et.~al.} \cite{XAI}, p.~193.
\newblock
  \href{http://dx.doi.org/10.1007/978-3-030-28954-6_10}{\doi{10.1007/978-3-030-28954-6_10}}.

\bibitem{Agarwal:2020fpt}
G.~Agarwal\hrefCMSnoop {}{ {et~al.}, ``{Explainable AI for ML jet taggers using
  expert variables and layerwise relevance propagation}'',} \textit{ JHEP}
  \textbf{ 05} (2021) 208,
  \href{http://dx.doi.org/10.1007/JHEP05(2021)208}{\doi{10.1007/JHEP05(2021)208}},
  \href{http://www.arXiv.org/abs/2011.13466}{\texttt{arXiv:2011.13466}}.

\bibitem{Lai:2020byl}
\hrefCMSnoop {}{Y.~S. Lai, D.~Neill, M.~P\l{}osko\'n, and F.~Ringer,
  ``{Explainable machine learning of the underlying physics of high-energy
  particle collisions}'',}
  \href{http://www.arXiv.org/abs/2012.06582}{\texttt{arXiv:2012.06582}}.

\bibitem{Neubauer:2022zpn}
\hrefCMSnoop {}{M.~S. Neubauer and A.~Roy, ``{Explainable AI for High Energy
  Physics}'',} in \textit{ {2022 Snowmass Summer Study}}.
\newblock 2022.
\newblock
  \href{http://www.arXiv.org/abs/2206.06632}{\texttt{arXiv:2206.06632}}.

\bibitem{Mokhtar:2021bkf}
F.~Mokhtar\href
  {https://ml4physicalsciences.github.io/2021/files/NeurIPS_ML4PS_2021_120.pdf}{
  {et~al.}, ``{Explaining machine-learned particle-flow reconstruction}'',} in
  \textit{ {4th Machine Learning and the Physical Sciences Workshop at the 35th
  Conference on Neural Information Processing Systems}}.
\newblock 2021.
\newblock
  \href{http://www.arXiv.org/abs/2111.12840}{\texttt{arXiv:2111.12840}}.

\bibitem{Pata:2021oez}
J.~Pata\hrefCMSnoop {}{ {et~al.}, ``{MLPF: Efficient machine-learned
  particle-flow reconstruction using graph neural networks}'',} \textit{ Eur.
  Phys. J. C} \textbf{ 81} (2021), no.~5, 381,
  \href{http://dx.doi.org/10.1140/epjc/s10052-021-09158-w}{\doi{10.1140/epjc/s10052-021-09158-w}},
  \href{http://www.arXiv.org/abs/2101.08578}{\texttt{arXiv:2101.08578}}.

\bibitem{Pata:2022wam}
J.~Pata\hrefCMSnoop {}{ {et~al.}, ``{Machine Learning for Particle Flow
  Reconstruction at CMS}'',} in \textit{ {20th International Workshop on
  Advanced Computing and Analysis Techniques in Physics Research}: {AI Decoded
  - Towards Sustainable, Diverse, Performant and Effective Scientific
  Computing}}.
\newblock 2022.
\newblock
  \href{http://www.arXiv.org/abs/2203.00330}{\texttt{arXiv:2203.00330}}.

\bibitem{GNNLRP}
T.~Schnake\hrefCMSnoop {}{ {et~al.}, ``Higher-order explanations of graph
  neural networks via relevant walks'',}
  \href{http://www.arXiv.org/abs/2006.03589}{\texttt{arXiv:2006.03589}}.

\bibitem{GLRP}
H.~Chereda\hrefCMSnoop {}{ {et~al.}, ``Explaining decisions of graph
  convolutional neural networks: patient-specific molecular subnetworks
  responsible for metastasis prediction in breast cancer'',} \textit{ Genome
  Medicine} \textbf{ 13} (2021), no.~1, 42,
  \href{http://dx.doi.org/10.1186/s13073-021-00845-7}{\doi{10.1186/s13073-021-00845-7}}.

\bibitem{Kasieczka:2019dbj}
\hrefCMSnoop {}{A.~Butter {et~al.}, ``{The Machine Learning landscape of top
  taggers}'',} \textit{ SciPost Phys.} \textbf{ 7} (2019) 014,
  \href{http://dx.doi.org/10.21468/SciPostPhys.7.1.014}{\doi{10.21468/SciPostPhys.7.1.014}},
  \href{http://www.arXiv.org/abs/1902.09914}{\texttt{arXiv:1902.09914}}.

\bibitem{kasieczka_gregor_2019_2603256}
\hrefCMSnoop {}{G.~Kasieczka, T.~Plehn, J.~Thompson, and M.~Russel, ``Top quark
  tagging reference dataset'',} 2019.
\newblock
  \href{http://dx.doi.org/10.5281/zenodo.2603256}{\doi{10.5281/zenodo.2603256}},
  \url {https://doi.org/10.5281/zenodo.2603256}.

\bibitem{xai4hep}
\hrefCMSnoop {}{F.~Mokhtar, R.~Kansal, and J.~Duarte, ``{xai4hep toolbox}'',}
  2022.
\newblock
  \href{http://dx.doi.org/10.5281/zenodo.7266537}{\doi{10.5281/zenodo.7266537}},
  \url {https://github.com/farakiko/xai4hep}.

\bibitem{Thaler:2010tr}
\hrefCMSnoop {}{J.~Thaler and K.~Van~Tilburg, ``{Identifying Boosted Objects
  with N-subjettiness}'',} \textit{ JHEP} \textbf{ 03} (2011) 015,
  \href{http://dx.doi.org/10.1007/JHEP03(2011)015}{\doi{10.1007/JHEP03(2011)015}},
  \href{http://www.arXiv.org/abs/1011.2268}{\texttt{arXiv:1011.2268}}.

\bibitem{Sjostrand:2014zea}
T.~Sj\"{o}strand\hrefCMSnoop {}{ {et~al.}, ``{An introduction to
  \PYTHIA8.2}'',} \textit{ Comput. Phys. Commun.} \textbf{ 191} (2015) 159,
  \href{http://dx.doi.org/10.1016/j.cpc.2015.01.024}{\doi{10.1016/j.cpc.2015.01.024}},
  \href{http://www.arXiv.org/abs/1410.3012}{\texttt{arXiv:1410.3012}}.

\bibitem{deFavereau:2013fsa}
\hrefCMSnoop {}{{DELPHES 3} Collaboration, ``{\DELPHES3}, a modular framework
  for fast simulation of a generic collider experiment'',} \textit{ JHEP}
  \textbf{ 02} (2014) 057,
  \href{http://dx.doi.org/10.1007/JHEP02(2014)057}{\doi{10.1007/JHEP02(2014)057}},
  \href{http://www.arXiv.org/abs/1307.6346}{\texttt{arXiv:1307.6346}}.

\bibitem{Cacciari:2008gp}
\hrefCMSnoop {}{M.~Cacciari, G.~P. Salam, and G.~Soyez, ``{The anti-$\kt$ jet
  clustering algorithm}'',} \textit{ JHEP} \textbf{ 04} (2008) 063,
  \href{http://dx.doi.org/10.1088/1126-6708/2008/04/063}{\doi{10.1088/1126-6708/2008/04/063}},
  \href{http://www.arXiv.org/abs/0802.1189}{\texttt{arXiv:0802.1189}}.

\bibitem{Cacciari:2011ma}
\hrefCMSnoop {}{M.~Cacciari, G.~P. Salam, and G.~Soyez, ``{\FastJet} user
  manual'',} \textit{ Eur. Phys. J. C} \textbf{ 72} (2012) 1896,
  \href{http://dx.doi.org/10.1140/epjc/s10052-012-1896-2}{\doi{10.1140/epjc/s10052-012-1896-2}},
\href{http://www.arXiv.org/abs/1111.6097}{\texttt{arXiv:1111.6097}}.
%%CITATION = ARXIV:1111.6097;%%.

\bibitem{PyTorchGeometric}
\href {https://pytorch-geometric.readthedocs.io/}{M.~Fey and J.~E. Lenssen,
  ``Fast graph representation learning with {PyTorch Geometric}'',} in \textit{
  ICLR Workshop on Representation Learning on Graphs and Manifolds}.
\newblock 2019.
\newblock
  \href{http://www.arXiv.org/abs/1903.02428}{\texttt{arXiv:1903.02428}}.

\bibitem{DGCNN}
Y.~Wang\hrefCMSnoop {}{ {et~al.}, ``Dynamic graph {CNN} for learning on point
  clouds'',} \textit{ ACM Trans. Graph.} \textbf{ 38} (2019)
  \href{http://dx.doi.org/10.1145/3326362}{\doi{10.1145/3326362}},
  \href{http://www.arXiv.org/abs/1801.07829}{\texttt{arXiv:1801.07829}}.

\bibitem{sanity_saliency}
J.~Adebayo\href
  {https://proceedings.neurips.cc/paper/2018/file/294a8ed24b1ad22ec2e7efea049b8737-Paper.pdf}{
  {et~al.}, ``Sanity checks for saliency maps'',} in \textit{ Advances in
  Neural Information Processing Systems}, S.~Bengio {et~al.}, eds., volume~31.
\newblock Curran Associates, Inc., 2018.
\newblock
  \href{http://www.arXiv.org/abs/1810.03292}{\texttt{arXiv:1810.03292}}.

\bibitem{Dokshitzer:1997in}
\hrefCMSnoop {}{Y.~L. Dokshitzer, G.~D. Leder, S.~Moretti, and B.~R. Webber,
  ``{Better jet clustering algorithms}'',} \textit{ JHEP} \textbf{ 08} (1997)
  001,
  \href{http://dx.doi.org/10.1088/1126-6708/1997/08/001}{\doi{10.1088/1126-6708/1997/08/001}},
  \href{http://www.arXiv.org/abs/hep-ph/9707323}{\texttt{arXiv:hep-ph/9707323}}.

\bibitem{Wobisch:1998wt}
\hrefCMSnoop {}{M.~Wobisch and T.~Wengler, ``{Hadronization corrections to jet
  cross-sections in deep inelastic scattering}'',} in \textit{ {Workshop on
  Monte Carlo Generators for HERA Physics (Plenary Starting Meeting)}}, p.~270.
\newblock 1998.
\newblock
  \href{http://www.arXiv.org/abs/hep-ph/9907280}{\texttt{arXiv:hep-ph/9907280}}.

\bibitem{CMS-PAS-JME-09-001}
\href {https://cds.cern.ch/record/1194489}{{CMS} Collaboration, ``{A
  Cambridge-Aachen (C-A) based Jet Algorithm for boosted top-jet tagging}'',}
  {CMS Physics Analysis Summary} CMS-PAS-JME-09-001, 2009.

\bibitem{Larkoski:2013eya}
\hrefCMSnoop {}{A.~J. Larkoski, G.~P. Salam, and J.~Thaler, ``{Energy
  Correlation Functions for Jet Substructure}'',} \textit{ JHEP} \textbf{ 06}
  (2013) 108,
  \href{http://dx.doi.org/10.1007/JHEP06(2013)108}{\doi{10.1007/JHEP06(2013)108}},
  \href{http://www.arXiv.org/abs/1305.0007}{\texttt{arXiv:1305.0007}}.

\bibitem{Moult:2016cvt}
\hrefCMSnoop {}{I.~Moult, L.~Necib, and J.~Thaler, ``{New Angles on Energy
  Correlation Functions}'',} \textit{ JHEP} \textbf{ 12} (2016) 153,
  \href{http://dx.doi.org/10.1007/JHEP12(2016)153}{\doi{10.1007/JHEP12(2016)153}},
  \href{http://www.arXiv.org/abs/1609.07483}{\texttt{arXiv:1609.07483}}.

\bibitem{Komiske:2017aww}
\hrefCMSnoop {}{P.~T. Komiske, E.~M. Metodiev, and J.~Thaler, ``{Energy flow
  polynomials: A complete linear basis for jet substructure}'',} \textit{ JHEP}
  \textbf{ 04} (2018) 013,
  \href{http://dx.doi.org/10.1007/JHEP04(2018)013}{\doi{10.1007/JHEP04(2018)013}},
  \href{http://www.arXiv.org/abs/1712.07124}{\texttt{arXiv:1712.07124}}.

\bibitem{interpretability_and_robustness}
\hrefCMSnoop {}{H.~Fokkema, R.~de~Heide, and T.~van Erven, ``Attribution-based
  explanations that provide recourse cannot be robust'',}
  \href{http://www.arXiv.org/abs/2205.15834}{\texttt{arXiv:2205.15834}}.

\end{thebibliography}\endgroup
\clearpage

\appendix

\section{ParticleNet training}
\label{app:performance}

The ParticleNet model is implemented in PyTorch Geometric. We choose $k=12$, instead of $k=16$ which was proposed by the ParticleNet paper, as the number of nearest neighbors to perform kNN. 
We also choose EdgeConv blocks with two (instead of three) fully-connected deep neural network linear layers, each followed by a batch normalization layer and a ReLU activation function.
We believe these changes will decrease the computation cost and should not change the conclusions.

\begin{figure}[H]
    % \centering
    % \includegraphics[width=0.5\textwidth]{figs/roc_curve.pdf}
    % \includegraphics[width=0.5\textwidth]{figs/loss.pdf}    
    \includegraphics[width=1\textwidth]{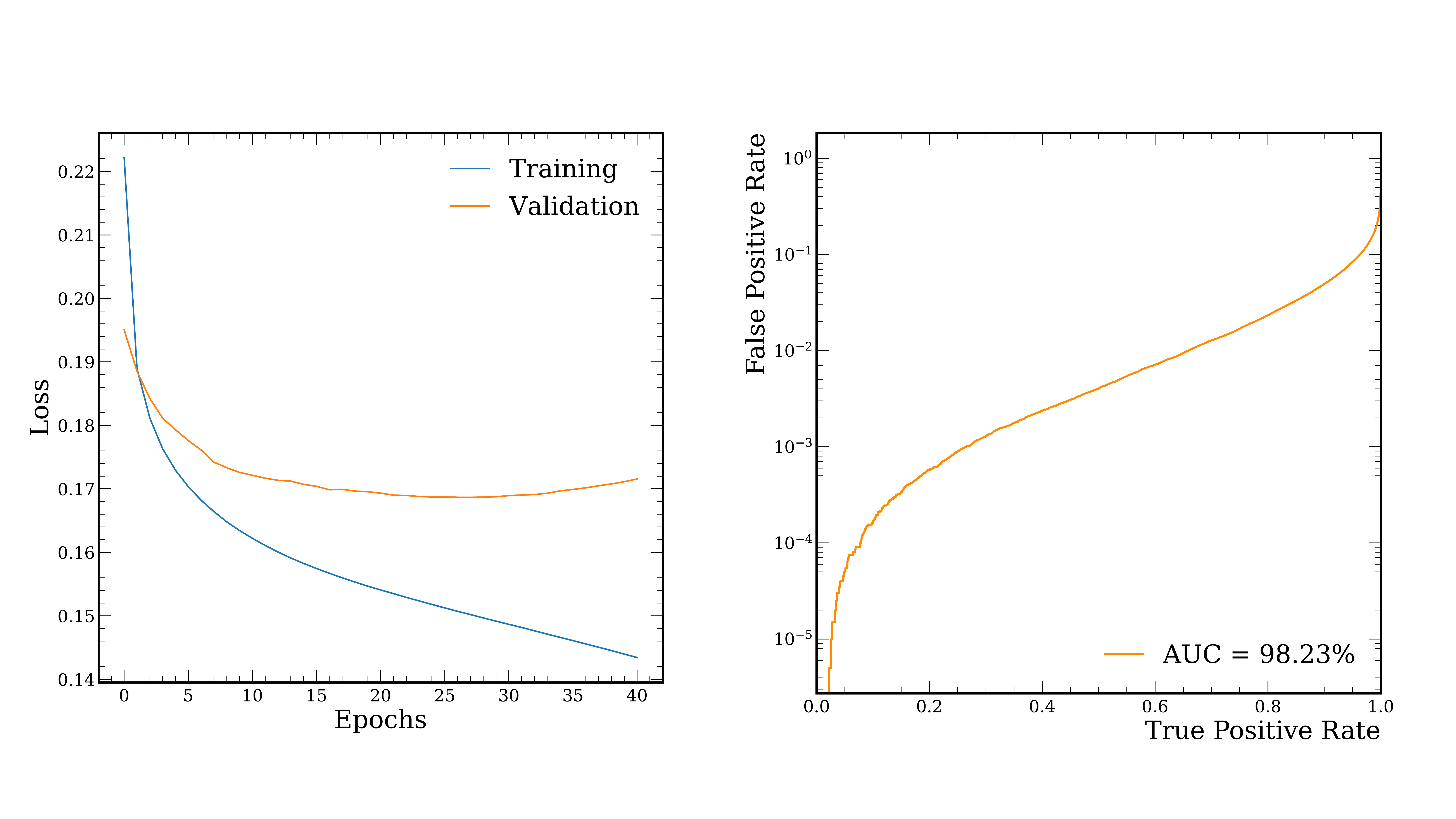}    
    \caption{
        Training and performance of the ParticleNet model. 
        A decreasing, convergent loss plot is presented (left) and the classification accuracy is presented in the form of a receiver operating characteristic (ROC) curve (right). 
        We can see that the model stops learning after 27 epochs.
    }
    \label{fig:performance}
\end{figure}

We perform the training over 1.2\,M events, with an additional 400\,k events used for validation, and 400\,k events used for testing. 
We run a training for 40 epochs with early stopping on a single Nvidia GTX 1080Ti graphics processing unit (GPU). 
The training converges after 27 epochs which takes about 3 hours. We use the Adam optimizer with a learning rate of $10^{-4}$.
The training performance is shown in \autoref{fig:performance}.

\clearpage

\section{Sample edge $R$ graphs}
\label{app:sample_graphs}

\begin{figure}[h!]
\centering
\includegraphics[width=0.94\textwidth]{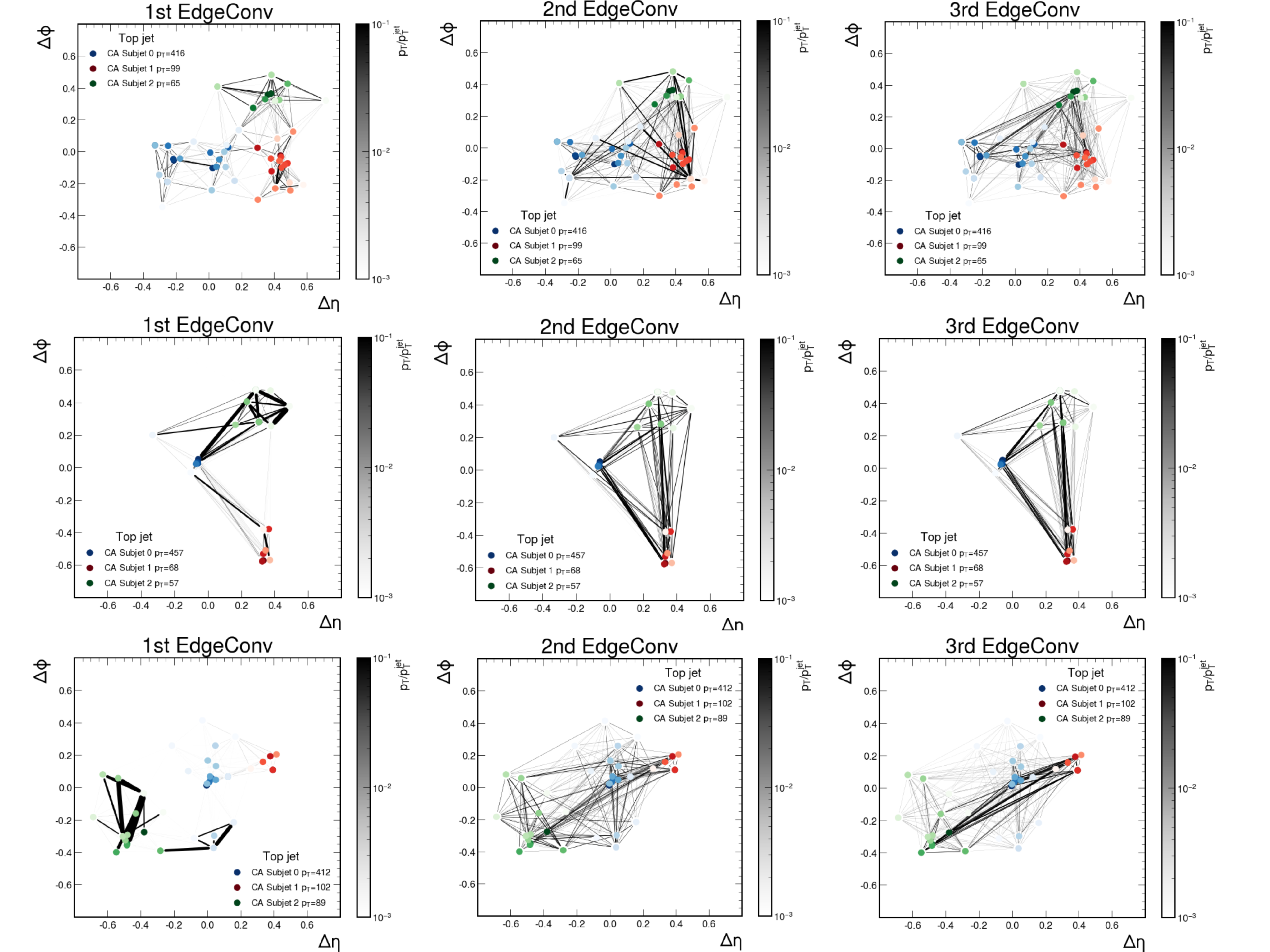}

\includegraphics[width=0.94\textwidth]{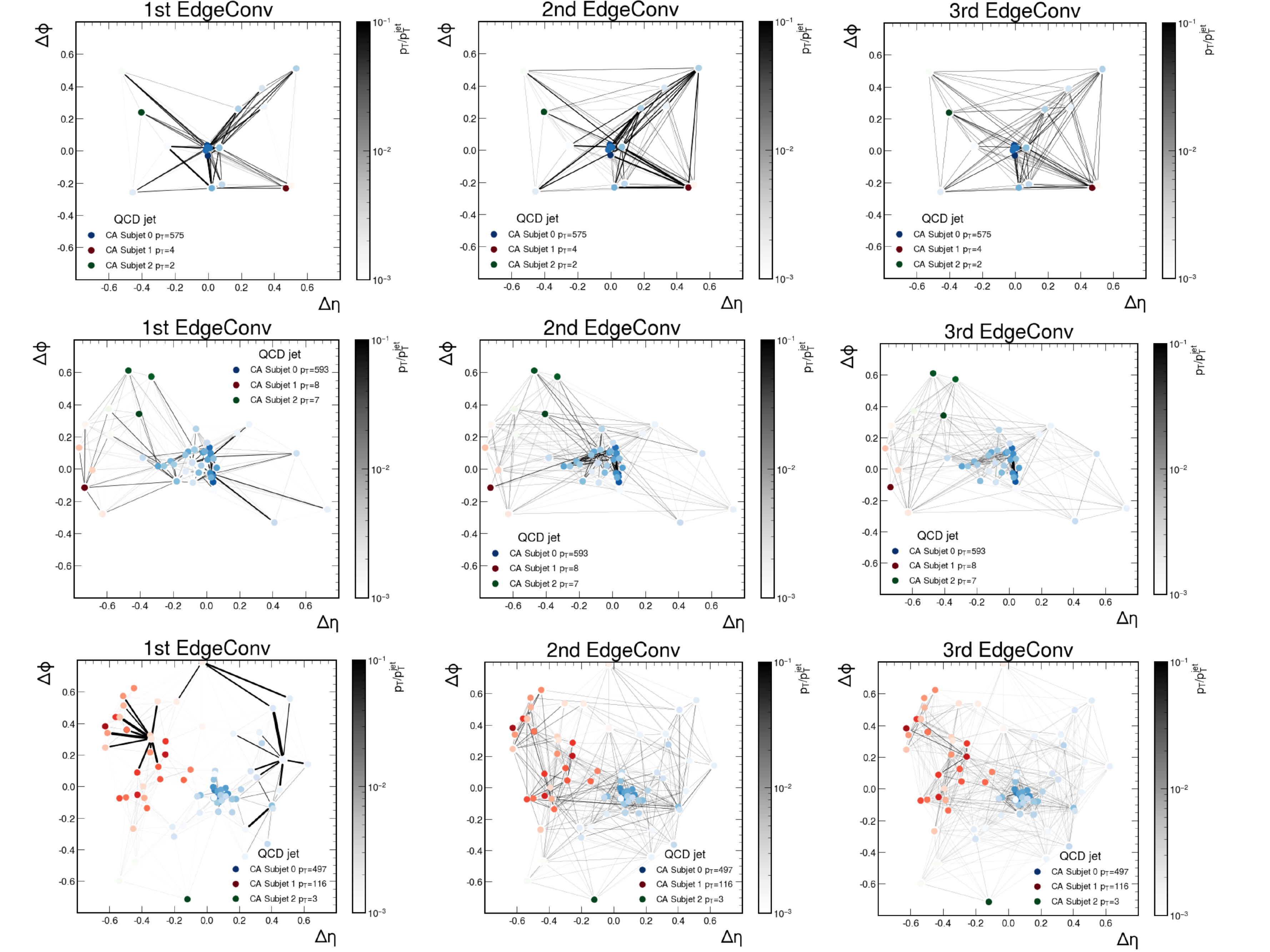}
\caption{Sample edge $R$ graphs for top quark (upper 9 graphs) and QCD jets (lower 9 graphs).}
\label{fig:samples}
\end{figure}

We present sample edge $R$ graphs for top quark (QCD) jets in Fig.~\ref{fig:samples} in the upper (lower) panel.
We observe that by the last EdgeConv block, the model is learning to connect nodes farther apart in $(\eta, \phi)$ space.

\clearpage

\end{document}